\documentclass[11pt]{article} 
\usepackage{hyperref} 
\pdfoutput=1
\begin{document} 
\title{Acoustically bound crystals}
\author{P. Marmottant \and  D. Rabaud \and P. Thibault \and  M. Mathieu \\ Laboratoire de Spectrom\'etrie Physique \\ÊCNRS and University of Grenoble, \\ Grenoble, France}
\maketitle
\begin{abstract} 
In these fluid dynamics videos, we show how bubbles flowing in a thin microchannel interact under an acoustic field. Because of acoustic interactions without direct contact, bubbles self-organize into periodic patterns, and spontaneously form acoustically bound crystals. We also present the interaction with boundaries, equivalent to the interaction with image bubbles, and unravel the peculiar vibration modes of the confined bubbles.
\end{abstract}

We generate microbubbles, with a diameter ranging from 20 to  50 micrometer,
 confined within thin 25 micrometers high elastomeric channels made of polydimethylsiloxane. 
The microbubbles are generated by flow-focusing a gas jet with a solution of surfactants.
The bubbles are highly confined in between the top and bottom walls, and have therefore the shape a squeezed sphere shape with two flat faces (white on images) and a curved perimeter (black on images).
We then apply an acoustic field by molding a glass plate into the elastomer , just above the microchannel, separated by only 145 micrometers. We set a standing wave in the glass rod, and the sound is emitted through the elastomer, that transmits efficiently sound to the channel.

We observe that bubbles vibrate and interact, with acoustic forces called secondary Bjerknes forces. This interaction has a specificity: it presents a minimum at a finite distance, and a minimum at contact. Bubbles either keep a fixed distance or either agglomerate without coalescing, because surfactants impede the contact of interfaces. Spontaneous patterns therefore occur.

The interaction is believed to be mediated by surface waves on the elastomer,  which is comforted by the fact that they have a very slow velocity and therefore a small wavelength comparable to the observed distances equilibrium distances (the wavelength of sound in water is 50 times larger).
Another fact comforting this scenario, is that it explains why bubbles are attracted to walls or keep a fixed distance. Indeed surface waves reflect on boundaries, which is equivalent to placing an image bubble symmetrically to the boundary. Bubbles then interact with their own images.

The vibration mode of the bubbles is not always an axisymmetric pulsation. At large sound amplitude we observe undulations, that are due to a surface instability at the free surface of the bubble. It creates standing waves on the perimeter. Large bubbles present a larger number of crests.

In conclusion, bubbles can form acoustically bound crystals, 
which are autonomous structures flowing with the liquid.

 \end{document}